\documentclass[fp,twocolumn]{jpsj3}
\usepackage{amsmath,amssymb}
\usepackage{mathrsfs}
\def\v#1{\mib #1}
\def\dfrac#1#2{{\displaystyle\frac{#1}{#2}}}

\newcommand{\aver}[1]{\left\langle {#1} \right\rangle}
\def\Jda{J_{\rm d1}}
\def\Jdb{J_{\rm d2}}
\def\Jd{J_{\rm d}}
\def\Jr{J_{\rm r}}
\def\Jl{J_{\rm l}}
\def\Jlc{{\Jl}_{\rm c}}
\def\Jinst{{\Jl}_{\rm s}}
\def\Sqra{S_{\rm intra}}
\def\Sqer{S_{\rm inter}}
\def\Sqerc{S_{\rm inter}^{\rm c}}
\def\Sqers{S_{\rm inter}^{\rm s}}

\def\simgeq{\mbox{\raisebox{-1.0ex}{$\stackrel{>}{\sim}$}}}

\title
{
Modified Spin Wave Analysis of Low Temperature Properties of Spin-1/2 Frustrated  Ferromagnetic Ladder}
\author
{
Kazuo {\sc Hida}\thanks{E-mail address: hida@mail.saitama-u.ac.jp} and  Takashi {\sc Iino}\thanks{deceased, June 2011}}

\inst
{Division of Material Science, Graduate School of Science and Engineering, \\ Saitama University, Saitama, Saitama, 338-8570\\ 
}

\recdate
{
November 28, 2011
}

\abst
{
Low temperature properties of the spin-1/2 frustrated ladder with ferromagnetic rungs and legs, and two different antiferromagnetic next nearest neighbor interaction are investigated using the modified spin wave approximation in the region with ferromagnetic ground state. The temperature dependence of the magnetic susceptibility and magnetic structure factors is calculated. The results are consistent with the numerical exact diagonalization results in the intermediate temperature range. Below this temperature range, the finite size effect is significant in the numerical diagonalization results, while the modified spin wave approximation gives more reliable results. The low temperature properties near the limit of the stability of the ferromagnetic ground state are also discussed.}

\kword
{frustration, ferromagnetic ladder, modified spin wave, magnetic susceptibility, magnetic structure factor
}

\begin{document}
\sloppy
\maketitle
\section{Introduction}

Among a variety of models and materials with strong quantum fluctuation, quantum spin ladders have been attracting the interest of many condensed matter physicists. Recently, a spin-1/2 ferromagnetic ladder compound {Cu$^{\rm II}$Cl($O$-$mi$)}$_2$($\mu$-Cl)$_2$ ($mi$ = 2-methylisothiazol-3(2$H$)-one) is synthesized with possible frustrated interrung interaction.\cite{kato.ejic10} 

Although frustrated and unfrustrated spin ladders with nonmagnetic ground states have been extensively studied\cite{Barnes.PRB93,Gopalan.PRB94,hik.prb10,HakobyanPRB01}, those with  ferromagnetic ground states have been less studied. This would be due to the simplicity of the ferromagnetic ground state. Even if the ground state is ferromagnetic, however, the excited states and  finite temperature properties should be influenced by frustration. Especially, near the stability limit of the ferromagnetic ground state, we expect characteristic  temperature dependence of physical quantities. In the present work, we investigate the finite temperature properties of frustrated spin ladders with ferromagnetic ground states using the modified spin wave (MSW) approximation\cite{taka.ptp86,Takahashi.PRL87,Takahashi.PRB90,Takahashi.PRB89,Ohara-YosidaJPSJ89,Ohara-YosidaJPSJ90} and the numerical exact diagonalization (ED) calculation for short ladders.

The MSW approximation was first proposed by Takahashi\cite{taka.ptp86} to investigate the low temperature properties of ferromagnetic chains.  If the conventional spin wave approximation is applied to one-dimensional ferromagnets, the thermal fluctuation of magnetization diverges at finite temperatures. This is natural, considering the absence of finite temperature long range order in one dimension\cite{Mermin.PRL66}. This divergence, however, prevents the calculation of  physical properties such as magnetic susceptibility and magnetic structure factor at finite temperatures. To circumvent this difficulty, Takahashi proposed to introduce the chemical potential $\mu$ for magnons and to fix $\mu$ by imposing the constraint that the expectation value of the total magnetization  $S^z_{\rm tot}$ vanishes as
\begin{align}
\aver{S^z_{\rm tot}}=0 \label{constraint}
\end{align}
taking account of the Mermin-Wagner theorem\cite{Mermin.PRL66}. This procedure is quite successful for one and two-dimensional ferromagnets\cite{taka.ptp86,Takahashi.PRL87,Takahashi.PRB90} and two-dimensional antiferromangets\cite{Takahashi.PRB89,Ohara-YosidaJPSJ89,Ohara-YosidaJPSJ90}.  Recently, this method has been successfully applied to dimerized ferromagnetic chains\cite{HerzogPRB11}. In the present work, we employ this method to calculate the low temperature magnetic susceptibility of the present model with ferromagnetic ground states. We also calculate the magnetic structure factor to confirm that the short range correlation is properly described within the MSW approximation.

\begin{figure}
%\centerline{\includegraphics[width=6cm]{ffladder_mono.eps}}
\centerline{\includegraphics[width=6cm]{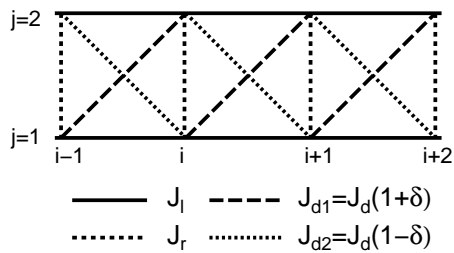}}
\caption{Structure of frustrated ladder studied in the present work.}
\label{lattice}
\end{figure}

This paper is organized as follows. In the next section, the model Hamiltonian is introduced. The MSW analysis  is explained  in \S 3. In \S 4, the MSW equations are numerically solved and the results are compared with the ED calculation for short chains. The last  section is devoted to summary and discussion. The low temperature expansion of the MSW equations is explained in Appendix.
\section{Hamiltonian}
\label{section:ham}

We consider the frustrated ferromagnetic ladder with spin-1/2 described by the Hamiltonian 
\begin{align}
{\cal H} &=\sum_{i=1}^{L} \left[\Jl(\v{S}_{i,1}\v{S}_{i+1,1}+\v{S}_{i,2}\v{S}_{i,2})\right.\nonumber\\
&\left.+\Jr\v{S}_{i,1}\v{S}_{i,2}+\Jda\v{S}_{i,1}\v{S}_{i+1,2}+\Jdb\v{S}_{i+1,1}\v{S}_{i,2}\right],\label{hama}
\end{align}
where $\v{S}_{i,j}$  $(i=1,...,L, j=1,2)$ are the spin operators with magnitude $S$. For the numerical calculation, we only consider the case of $S=1/2$. The number of the unit cells is denoted by $L$. In this paper, we focus on the case $\Jl, \Jr < 0$ (ferromagnetic) and $\Jda, \Jdb > 0$ (antiferromagnetic). The lattice structure is schematically depicted in Fig. \ref{lattice}. We also use the parameterization
\begin{align}
\Jda&=\Jd(1+\delta),\ \Jdb=\Jd(1-\delta)
\end{align}
if convenient.
\section{Modified Spin Wave Analysis}
\subsection{Formulation}

We employ the standard Holstein-Primakoff transformation, 
\begin{align}
\begin{split}
  &S_{i,j}^+=S_{i,j}^x+iS_{i,j}^y=\sqrt{2S}f_{i,j}(S)a_{i,j}, \\
  &S_{i,j}^-=S_{i,j}^x-iS_{i,j}^y=\sqrt{2S}a_{i,j}^{\dagger }f_{i,j}(S), \\
  &S_{i,j}^z=S-a_{i,j}^{\dagger }a_{i,j}, \\
  &f_{i,j}(S)=\sqrt{1-(2S)^{-1}a_{i,j}^{\dagger }a_{i,j}}%=1-(4S)^{-1}a_{i,j}^{\dagger }a_{i,j}-O(S^{-2})
\ (i=1,...,L; j=1,2), \label{HP}
\end{split}
\end{align}
where $a_{i,j}^{\dagger }$ and $a_{i,j}$ are magnon creation and annihilation operators on the $i$-th rung and the $j$-th leg. After the Fourier transformation with respect to $i$, the Hamiltonian is rewritten as
\begin{align}
{\cal H}&={\cal H}_0+{\cal H}_1,\\
{\cal H}_0&= \sum_k\left( a_{k,1}^{\dagger } \: a_{k,2}^{\dagger } \right) %\nonumber\\ \times&
\left( \begin{array}{cc} E_1  & E_2 \\
                      %& \\
 E_2^* & E_1 \\
\end{array} \right)\left( \begin{array}{c} a_{k,1} \\  %\\ 
a_{k,2} 
\end{array} \right),\\
 {\cal H}_1 &=-\frac{\Jl}{4L}\sum _{kk^{\prime }q}\sum _{j=1}^2\left\{ e^{-i(k^{\prime }-q)}(1-e^{ik^{\prime }})(1-e^{ik}) \right. \notag \\
           &\left. +e^{ik}(1-e^{-i(k+q)})(1-e^{-i(k^{\prime }-q)})\right\}\nonumber\\
&\times a_{k+q,j}^{\dagger }a_{k^{\prime }-q,j}^{\dagger }a_{k^{\prime },j}a_{k,j} \nonumber\\
   &-\frac{\Jr}{4L}\sum _{kk^{\prime }q}\left\{ a_{k+q,1}^{\dagger }a_{k^{\prime }-q,2}^{\dagger }\right.\nonumber\\
&\left.\times(a_{k,1}-a_{k,2})(a_{k^{\prime },1}-a_{k^{\prime },2})
    +\text{h.c}\right\} \nonumber\\
         &-\frac{\Jda}{4L}\sum _{kk^{\prime }q}\left\{ e^{-i(k^{\prime }-q)}a_{k+q,1}^{\dagger }a_{k^{\prime }-q,2}^{\dagger }\right.\nonumber\\
&\left.\times(a_{k,1}-e^{ik}a_{k,2})(a_{k^{\prime },1}-e^{ik^{\prime }}a_{k^{\prime },2})+\text{h.c} \right\} \nonumber\\
           &-\frac{\Jdb}{4L}\sum _{kk^{\prime }q}\left\{ e^{-i(k^{\prime }-q)}a_{k+q,2}^{\dagger }a_{k^{\prime }-q,1}^{\dagger }\right.\nonumber\\
&\left.
          \times(a_{k,2}-e^{ik}a_{k,1})(a_{k^{\prime },2}-e^{ik^{\prime }}a_{k^{\prime },1})+\text{h.c} \right\} 
\end{align}
up to the quartic order in $a_{k,i}$ and $a_{k,i}^{\dagger }$. 
 Here, $E_1$ and $E_2$ are defined by
\begin{align}
\begin{split}
  E_1\equiv &\left\{ -2\Jl(1-\cos k)-\Jr-\Jda-\Jdb\right\} S, \\
  E_2\equiv &(\Jr+e^{ik}\Jda+e^{-ik}\Jdb)S.
\end{split}
\end{align}

We introduce the unitary transformation
\begin{align}
\begin{split}
  \alpha _k^{\dagger }&\equiv \frac{1}{\sqrt{2}}\left( a_{k1}^{\dagger }e^{i\frac{\phi (k)}{2}}+a_{k2}^{\dagger }e^{-i\frac{\phi (k)}{2}}\right),\\ 
  \beta _k^{\dagger }&\equiv \frac{1}{\sqrt{2}}\left( a_{k1}^{\dagger }e^{i\frac{\phi (k)}{2}}-a_{k2}^{\dagger }e^{-i\frac{\phi (k)}{2}}\right), \label{eq:uni}
\end{split}
\end{align}
and the corresponding magnon number operators
\begin{align} 
{\hat{n}_{\alpha k}}=  \alpha _k^{\dagger } \alpha _k, \qquad {\hat{n}_{\beta k}}=  \beta _k^{\dagger } \beta _k.
\end{align}

The phase $\phi (k)$ is determined afterwards to minimize the free energy. We assume the density matrix for the magnons in the form of a product of single-magnon density matrices as
\begin{align}
  \rho =\sum _{\{ n_{\alpha k},n_{\beta k}\}}\prod _kP_{\alpha k}(n_{\alpha k})P_{\beta k}(n_{\beta k})
    \big| \{ n_{\alpha k},n_{\beta k}\}\big\rangle \big\langle \{ n_{\alpha k},n_{\beta k}\}\big|,
\label{eq:dm}
\end{align}
where
\begin{align}
  |\{ n_{\alpha k},n_{\beta k}\} \rangle =\prod _k(n_{\alpha k}!n_{\beta k}!)^{-\frac{1}{2}}(\alpha _k^{\dagger })^{n_{\alpha k}}(\beta _k^{\dagger })^{n_{\beta k}}|0\rangle
\end{align}
is the eigenstate of ${\hat{n}_{\alpha k}}$ and ${\hat{n}_{\beta k}}$ with eigenvalues ${{n}_{\alpha k}}$ and ${{n}_{\beta k}}$. In the following, the expectation value with respect to (\ref{eq:dm}) is denoted by $\aver{...}$. The probability that the state with wave number $k$ is occupied by $n$ bosons is denoted by $P_{\nu k}(n)\ (\nu=\alpha$ or $\beta$). Therefore, the following normalization conditions are required for all $k$.
\begin{align}
  \sum _{n=0}^{\infty }P_{\alpha k}(n)=1 \; \; ,\; \; \sum _{n=0}^{\infty }P_{\beta k}(n)=1 \label{condition1}.
\end{align}
Accordingly, the free energy is given by
\begin{align}
F=\aver{{\cal H}_0}+\aver{{\cal H}_1}%\sum _{\nu=\alpha ,\beta }\left( \sum _k\varepsilon _{\nu}(k)\sum _{n=0}^{\infty }nP_{\nu k}(n)
+T\sum _{\nu=\alpha,\beta}\sum _k\sum _{n=0}^{\infty }P_{\nu k}(n)\ln P_{\nu k}(n)%\right), 
\label{FE}
\end{align}
where the expectation values $\aver{{\cal H}_0}$ and $\aver{{\cal H}_1}$ are expressed as follows
\begin{align}
  \aver{{\cal H}_0}&=\sum _k\left\{E_1\tilde{n}_{k}+\frac{1}{2}(E_2e^{-i\phi(k)}+E_2^*e^{i\phi(k)})\delta\tilde{n}_{k}\right\},\\
%\end{align}
%\begin{align}
  \aver{{\cal H}_1}&=
\frac{\Jl}{2L}\left( \sum _k(1-\cos{k})\tilde{n}_{k}\right) ^2 \nonumber\\
&+\frac{\Jr}{4L}\Bigg[ \sum _k\left(\tilde{n}_{k}
-\delta\tilde{n}_{\alpha k}\cos{\phi (k)}\right) \Bigg] ^2\nonumber \\
&+\frac{\Jda}{4L}\Bigg[ \sum _k\left( \tilde{n}_{k}
                                                            -\delta\tilde{n}_{k}\cos{(\phi (k)-k)}\right) \Bigg] ^2 \nonumber\\
&+\frac{\Jdb}{4L}\Bigg[ \sum _k\left( \tilde{n}_{k}
                                                            -\delta\tilde{n}_{k}\cos{(\phi (k)+k)}\right) \Bigg] ^2,
\end{align}
where
\begin{align}
\tilde{n}_{\nu k}&=\aver{{n}_{\nu k}}=\sum_{n}n P_{\nu k}(n) \ \ (\nu=\alpha, \beta).
\end{align}
We also define
\begin{align}
\tilde{n}_{k}&=\tilde{n}_{\alpha k}+\tilde{n}_{\beta k},\ \ \delta\tilde{n}_{k}=\tilde{n}_{\alpha k}-\tilde{n}_{\beta k}
\end{align}
for convenience. 
The condition (\ref{constraint}) reduces to
\begin{align}
  S=\frac{1}{2L}\sum _k\tilde{n}_{ k}.  \label{condition2}
\end{align}
Introducing the Lagrangian multipliers $\mu _{\nu k}$ and $\mu$ which account for the constraint (\ref{condition1}) and (\ref{condition2}), we minimize the following quantity $W$ with  respect to $P_{\nu k}$ and $\phi(k)$.
\begin{align}
  W=F-\sum _{\nu=\alpha ,\beta }\sum _k\mu _{\nu k}\sum _{n=0}^{\infty }P_{\nu k}(n)-\mu \sum _{\nu=\alpha ,\beta }\sum _k\sum _{n=0}^{\infty }nP_{\nu k}(n).
\end{align}
\subsection{Spin-Spin Correlation Function and Magnetic Susceptibility}

The spin-spin correlation function is expressed in terms of the magnon occupation numbers as
\begin{align}
  \langle \mbox{\boldmath $S$}_{i1}\mbox{\boldmath $S$}_{i^{\prime }1}\rangle
   =&\langle \mbox{\boldmath $S$}_{i2}\mbox{\boldmath $S$}_{i^{\prime }2}\rangle
   =\left( \frac{1}{2L}\sum _k\cos k(x_{i^{\prime }}-x_i)\tilde{n}_{k}\right) ^2, \label{eq:cor11}\\
  \langle \mbox{\boldmath $S$}_{i1}\mbox{\boldmath $S$}_{i^{\prime }2}\rangle
   =&\left( \frac{1}{2L}\sum _k\cos \left( \phi (k)-k(x_{i^{\prime }}-x_i)\right)\delta\tilde{n}_{k}\right) ^2,\label{eq:cor12} %\\
\end{align}
where $x_i$ is the position of the $i$-th site. Using these expressions, the magnetic susceptibility per spin $\chi $ is expressed as
\begin{align}
  \chi =&\frac{(g\mu_{\rm B})^2}{2LT}\sum _{ii^{\prime }}\sum _{j=1}^2\sum _{j'=1}^2 \langle S_{ij}^zS_{i^{\prime }j'}^z\rangle  \notag \\
 =&\frac{(g\mu_{\rm B})^2}{6LT}\sum _{ii^{\prime }}\sum _{j=1}^2\sum _{j'=1}^2\langle \v{S}_{ij}\cdot\v{S}_{i^{\prime }j'}\rangle  \notag \\
&=\frac{(g\mu_{\rm B})^2}{6LT}\sum _k(\tilde{n}_{\alpha k}^2+\tilde{n}_{\beta k}^2+\tilde{n}_{\alpha k}+\tilde{n}_{\beta k}), \label{chi}
\end{align}
where $\mu_{\rm B}$ is the Bohr magneton and $g$ is the gyromagnetic ratio.
\subsection{Magnetic Structure Factor}
We define the intraleg and interleg magnetic structure factors as
\begin{align}
\Sqra(q)&=\frac{1}{L}\sum_{i}\aver{\v{S}_{0,1}\v{S}_{i,1}}\exp(i x_i q)\notag\\
&=\frac{1}{L}\sum_{i}\aver{\v{S}_{0,2}\v{S}_{i,2}}\exp(i x_i q),\label{eq:sq1}\\
\Sqer(q)&=\frac{1}{L}\sum_{ i}\aver{\v{S}_{0,1}\v{S}_{ i,2}}\exp (i x_i q).\label{eq:sq2}
\end{align}
Substituting (\ref{eq:cor11}) and (\ref{eq:cor12})
 into (\ref{eq:sq1}) and (\ref{eq:sq2}),
we find
\begin{align}
\Sqra(q)&=S+\frac{1}{4L}\sum _k\tilde{n}_{k+q/2} \tilde{n}_{k-q/2}, \\
\Sqer(q)&=\Sqerc(q)+i\Sqers(q),\\
\Sqerc(q)&=\frac{1}{4L}\sum _k\cos \left(\phi (k+q/2)-\phi (k-q/2)\right)\notag\\
&\times\delta \tilde{n}_{k+q/2} \delta \tilde{n}_{k-q/2},\\
\Sqers(q)&=\frac{1}{4L}\sum _k\sin \left(\phi (k+q/2)-\phi (k-q/2)\right)\notag\\
&\times\delta \tilde{n}_{k+q/2} \delta \tilde{n}_{k-q/2}. 
\end{align}
It should be noted that $\Sqer(q)$ has an imaginary part because of the absence of space inversion symmetry $x \leftrightarrow -x$.

\subsection{Lowest order approximation}

To the lowest order approximation, we only consider the Hamiltonian ${\cal H}_0$ and impose the constraint (\ref{constraint}). In the following, we call this approximation the MSW0 approximation. Minimizing 
\begin{align}
  W_0&=\aver{{\cal H}_0}-T{\cal S}-\sum _{\nu=\alpha ,\beta }\sum _k\mu _{\nu k}\sum _{n=0}^{\infty }P_{\nu k}(n)\notag\\
&-\mu \sum _{\nu=\alpha ,\beta }\sum _k\sum _{n=0}^{\infty }nP_{\nu k}(n)
\end{align}
with respect to $\phi$, we find 
\begin{align}
   \frac{\partial W_0}{\partial \phi(k)}&=\frac{\partial \aver{{\cal H}_0}}{\partial \phi(k)}=-\frac{i}{2}(E_2e^{-i\phi(k)}-E_2^*e^{i\phi(k)})\delta\tilde{n}_{k}=0.
\end{align}
This leads to
\begin{align}
\Jr\sin\phi(k)+\Jda\sin(\phi(k)-k)+\Jdb\sin(\phi(k)+k)=0,\label{sce0_1}
\end{align}
which determines the phase $\phi(k)$ as
\begin{align}
  \tan{\phi (k)}=\frac{(\Jda-\Jdb)\sin{k}}{\Jr+(\Jda+\Jdb)\cos{k}}\ \ (-\pi/2 < \phi < \pi/2).\label{eq:phi}
\end{align}

Minimizing $W_0$ with respect to $P_{\nu k}$, we find
\begin{align}
 \frac{\partial W_0}{\partial P_{\nu k}}&= n\varepsilon_{\nu}(k)%\frac{\partial \aver{{\cal H}_0}}{\partial P_{\nu k}}
+T(\ln P_{\nu k}+1)-n\mu-\mu_{\nu k}=0 \ \ (\nu=\alpha, \beta),
\label{sce0_2}%\\
\end{align}
where
\begin{align}
 \varepsilon_{\nu}(k)=\frac{1}{n}\frac{\partial \aver{{\cal H}_0}}{\partial P_{\nu k}} %,\ \frac{\partial \aver{{\cal H}_0}}{\partial P_{\beta k}}&= n  \varepsilon_{\beta}(k)
\end{align}
are  two branches of the magnon excitation energy given by
\begin{align}
 \varepsilon_{\alpha\atop\beta}(k)&= E_1\pm\frac{1}{2}(E_2e^{-i\phi(k)}+E_2^*e^{i\phi(k)})\nonumber\\
&= S \left\{-2\Jl(1-\cos k)-\Jr(1\mp\cos \phi(k))\right.\nonumber\\
&\left.-\Jda(1\mp\cos(k-\phi(k))-\Jdb(1\mp\cos(k+\phi(k))\right\}.%\\
\end{align}
The Hamiltonian ${\cal H}_0$ is rewritten as
\begin{align}
{\cal H}_0=&\sum_k  \begin{pmatrix}
            \alpha _k^{\dagger } & \beta _k^{\dagger }
          \end{pmatrix}
          \begin{pmatrix}
            \varepsilon _\alpha(k) & 0 \\ 0 & \varepsilon _\beta(k) 
          \end{pmatrix}
          \begin{pmatrix}
            \alpha _k & \beta _k
          \end{pmatrix}.
\end{align}
Eliminating $\mu_{\nu k}$ from (\ref{sce0_2}) %and (\ref{sce0_3})
 using the condition (\ref{condition1}), we find
\begin{align}
  P_{\nu k}(n)=\left\{ 1-e^{-(\varepsilon_{\nu}(k)-\mu )/T}\right\} e^{-n(\varepsilon _{\nu}(k)-\mu )/T}.\label{prob0}
\end{align}
The expectation value of $n_{\nu k}$ %and $n_{\nu k}^2$ 
is given by
\begin{align}
  \tilde{n}_{\nu k} &=\sum _{n=0}^{\infty }nP_{\nu k}(n)=%\left\{ 1-e^{-(\varepsilon_{\nu}(k)-\mu )/T}\right\} \sum _{n=0}^{\infty }ne^{-n(\varepsilon _{\nu}(k)-\mu )/T} \notag \\
\frac{1}{e^{(\varepsilon _{\nu}(k)-\mu )/T}-1}.\label{occu0}
\end{align}
The chemical potential $\mu$ is determined using the condition (\ref{condition2}).

\subsection{$O(S^0)$ approximation}

In this approximation, we include the terms of $O(S^0)$ in the Hamiltonian i.e.  ${\cal H}_1$. We call this approximation the MSW1 approximation. Minimizing
\begin{align}
  W_1&=\aver{{\cal H}_0}+\aver{{\cal H}_1}-T{\cal S}-\sum _{\nu=\alpha ,\beta }\sum _k\mu _{\nu k}\sum _{n=0}^{\infty }P_{\nu k}(n)\notag\\
&-\mu \sum _{\nu=\alpha ,\beta }\sum _k\sum _{n=0}^{\infty }nP_{\nu k}(n)
\end{align}
with respect to $\phi(k)$, we find
\begin{align}
   \frac{\partial W_1}{\partial \phi(k)}&=\frac{\partial \aver{{\cal H}_0}}{\partial \phi(k)}+\frac{\partial \aver{{\cal H}_1}}{\partial \phi(k)}=0%=\Bigg[-\frac{i}{2}(E_2e^{-i\phi(k)}-E_2^*e^{i\phi(k)}),
\end{align}
which yields
\begin{align}
&{\Jr}\sin{\phi (k)}S'_2+{\Jda}\sin{(\phi (k)-k)}S'_3\notag\\
&+{\Jdb}\sin{(\phi (k)+k)}S'_4=0,\label{sce1_1}
\end{align}
where
\begin{align}
S'_1&=\frac{1}{2L}\sum_{k}\cos {k}\tilde{n}_{k},\nonumber\\
S'_2&=\frac{1}{2L}\sum_{k}\cos (\phi (k))\delta\tilde{n}_{k},\nonumber\\
S'_3&=\frac{1}{2L}\sum_{k}\cos (\phi (k)-k)\delta\tilde{n}_{k},\nonumber\\
S'_4&=\frac{1}{2L}\sum_{k}\cos (\phi (k)+k)\delta\tilde{n}_{k}.
\end{align}
From eq. (\ref{sce1_1}), $\phi(k)$ is given as
\begin{align}
\tan{\phi (k)}&=\frac{({\Jda} S'_3-{\Jdb} S'_4)\sin k}{({\Jr}S'_2+({\Jda} S'_3+{\Jdb}S'_4)\cos k)}.
\end{align}
 Minimizing $W_1$ with respect to $P_{\nu k}$, we find
\begin{align}
  \frac{\partial W_1}{\partial P_{\nu k}}&= n\tilde{\varepsilon}_{\nu k}%\frac{\partial \aver{{\cal H}_0}}{\partial P_{\nu k}}+\frac{\partial \aver{{\cal H}_1}}{\partial P_{\nu k}}
+T(\ln P_{\nu k}+1)-n\mu-\mu_{\nu k}=0,
\label{sce1_2}
\end{align}
where
\begin{align}
   \tilde{\varepsilon}_{\nu k}&=\frac{1}{n}\frac{\partial }{\partial P_{\nu k}(n) }(\aver{{\cal H}_0}+\aver{{\cal H}_1})%, \\ 
\end{align}
is the dressed single magnon excitation energy given by
\begin{align}
&\tilde{\varepsilon}_{\alpha\atop\beta}(k)
= -2\Jl(1-\cos {k})S'_1-{\Jr}(1\mp\cos (\phi (k))S'_2\nonumber\\
&-{\Jda}(1\mp\cos (\phi (k)-k))S'_3-{\Jdb}(1\mp\cos (\phi (k)+k))S'_4.% \\
%\tilde{\varepsilon}_{\beta}(k)&=  -2\Jl(1-\cos {k})S'_1-{\Jr}(1+\cos (\phi (k))S'_2\nonumber\\
%&-{\Jda}(1+\cos (\phi (k)-k))S'_3-{\Jdb}(1+\cos (\phi (k)+k))S'_4. 
\end{align}
Expressions for  $P_{\nu k}(n)$ and $\tilde{n}_{\nu k}$ are obtained by replacing ${\varepsilon}_{\nu k}$ by $\tilde{\varepsilon}_{\nu k}$ in (\ref{prob0}) and (\ref{occu0}), respectively.
The phase $\phi(k)$ and chemical potential $\mu$ are determined by solving (\ref{sce1_1}) under the condition (\ref{condition2}) numerically.
\subsection{Low Temperature Behavior}

We employ the MSW0 approximation to investigate the low temperature behavior. Among two branches of the magnon excitation, $\varepsilon_{\beta}(k)$ has an energy gap at $k=0$. Therefore, we only consider $\varepsilon_{\alpha}(k)$ at low temperatures. Near $k=0$, the dispersion relation is given by
\begin{align}
  \varepsilon _\alpha(k) \simeq %& \Jr Sk^2\left[\jl -\frac{1-2\jd(1-\alpha^2)}{1-2\jd}\jd \right] \notag \\
                      %\equiv 
&\mathscr{J}Sk^2,
\end{align}
where
\begin{align}
  \mathscr{J} \equiv -\left[\Jl +\frac{\Jr(\Jda+\Jdb)+4\Jda\Jdb}{2(\Jr+\Jda+\Jdb)} \right]
\end{align}
is the effective ferromagnetic exchange interaction between the ferromagnetic rung dimers consisting of $\v{S}_{i,1}$ and $\v{S}_{i,2}$. For negative (ferromagnetic) $\Jl$ and $\Jr$, $\mathscr{J} $ is positive as long as the frustrating antiferromagnetic interaction $\Jda$ and $\Jdb$ are small. However, with the increase of $\Jda$ and $\Jdb$, $\mathscr{J}$ decreases and vanishes at 
\begin{align}
   \Jinst =-\frac{\Jr(\Jda+\Jdb)+4\Jda\Jdb}{2(\Jr+\Jda+\Jdb)}=-\frac{\Jr+2\Jd(1-\delta^2)}{\Jr+2\Jd}\Jd. \label{insta} 
\end{align}
This point is the limit of the stability of the ferromagnetic ground state. 

For $\mathscr{J}>0$,  the low temperature expansion can be carried out in the same manner as the ferromagnetic chain. The details are explained in Appendix. As a result, we obtain 
\begin{align}
  \chi \simeq &\frac{8(g\mu _{\text{B}})^2S^4\mathscr{J}}{3T^2}
              \left( 1-\frac{3}{4S}\frac{\zeta \! \left( \frac{1}{2}\right) }{\sqrt{\pi }}\sqrt{\frac{T}{\mathscr{J}S}}
                     +\frac{3}{16S^2}\frac{\zeta ^2\! \left( \frac{1}{2}\right) }{\pi }\frac{T}{\mathscr{J}S}\right).
\end{align}
Namely, the susceptibility is proportional to $T^{-2}$ at low temperatures. Defining $\tilde{\mathscr{J}}\equiv \mathscr{J}/2,S_{\text{eff}}\equiv 2S$, the susceptibility per rung is given by
\begin{align}
  &\chi_{\rm rung}=2\chi \notag\\
&\simeq \frac{2(g\mu _{\text{B}})^2S^4_{\text{eff}}\tilde{\mathscr{J}}}{3T^2}\left( 1-%\frac{3}{%S_{\text{eff}}}
\frac{3\zeta \left(\frac{1}{2}\right)}{\sqrt{2\pi }}
                 \sqrt{\frac{T}{2\tilde{\mathscr{J}}S^3_{\text{eff}}}}+%\frac{3}{S^2_{\text{eff}}}
                   \frac{3\zeta ^2\left(\frac{1}{2}\right)}{2\pi }\frac{T}{2\tilde{\mathscr{J}}S^3_{\text{eff}}}\right).
\end{align}
This coincides with the low temperature susceptibility of the ferromagnetic chain with spin $S_{\rm eff}$ and effective exchange interaction $\tilde{\mathscr{J}}$.\cite{taka.ptp86}

\section{Numerical Results}
\subsection{Results for the parameter set corresponding to Cu$^{\rm II}$Cl($O$-$mi$)$_2$($\mu$-Cl)$_2$}
\begin{figure}
%\centerline{\includegraphics[width=6cm]{frulad_chiT2g2.13_j_2.eps}}
\centerline{\includegraphics[width=6cm]{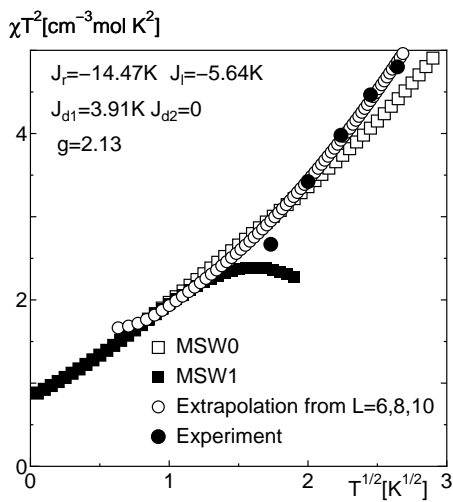}}
\caption{Magnetic susceptibility for exchange parameters corresponding to [{Cu$^{\rm II}$Cl($O$-$mi$)}$_2$($\mu$-Cl)$_2$] calculated by MSW0 (open squares) and MSW1 (filled squares) approximations. Open circles are Shanks extrapolation from the ED data for $L=6, 8$ and 10. The filled circles are experimental data taken from  ref. \citen{kato.ejic10}.
}
\label{sus}
\end{figure}

The magnetic susceptibility is calculated for the parameter set determined in ref. \citen{kato.ejic10} for Cu$^{\rm II}$Cl($O$-$mi$)$_2$($\mu$-Cl)$_2$ as shown in Fig. \ref{sus}.  The susceptibility calculated by  two MSW approximations and the Shanks extrapolation\cite{Shanks} from the ED data for $L=6, 8$ and 10 are shown. The experimental data for Cu$^{\rm II}$Cl($O$-$mi$)$_2$($\mu$-Cl)$_2$ are also shown. Although Fig. \ref{sus} contains only a limited number of experimental data, there are many data points  at higher temperatures and the exchange constants are determined by fitting them with the ED calculation as explained in ref. \citen{kato.ejic10}.

The MSW0 approximation reproduces the overall temperature dependence including relatively higher temperature regime. At low temperatures, where the ED results are strongly size dependent, both MSW approximations give the consistent results. The result of the MSW1 approximation is more smoothly connected to the ED data. However, it strongly deviates from the ED data as the temperature is raised. Actually, the MSW1 equations  have no solution other than $\phi=0, S'_1=S'_2=S'_3=0$ for $T \geq T^* \simeq 4K.$  
This implies a phase transition at  $T = T^*$. However, this  transition is spurious because the present system is one-dimensional and no phase transition should take place at finite temperatures.  Therefore, the MSW1 approximation is unreliable at $T \simgeq O(T^*)$. This spurious transition takes place even in the absence of frustration within the MSW1 approximation. A similar kind of spurious transition takes place in the square lattice Heisenberg antiferromagnet\cite{Takahashi.PRB89}. Hence, this is an artifact of the MSW1 approximation rather than the effect of frustration. On the contrary, the MSW0 approximation predicts no phase transition and gives the results qualitatively consistent with the ED results up to relatively high temperatures. 
\begin{figure}
%\centerline{\includegraphics[width=6cm]{mi_t2a.eps}}
%\centerline{\includegraphics[width=6cm]{mi_t3a.eps}}
%\centerline{\includegraphics[width=6cm]{mi_t4a.eps}}
\centerline{\includegraphics[width=6cm]{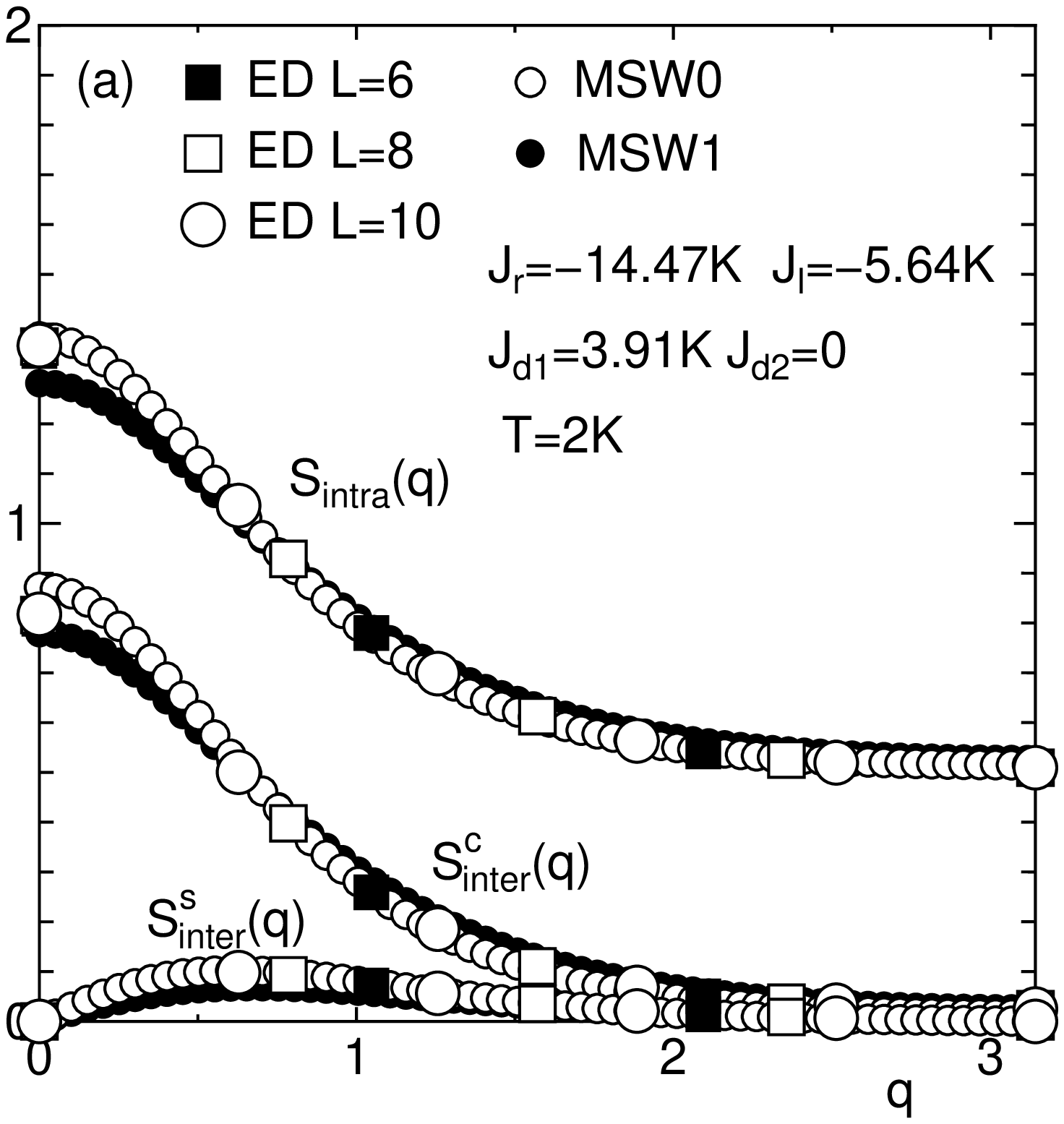}}
\centerline{\includegraphics[width=6cm]{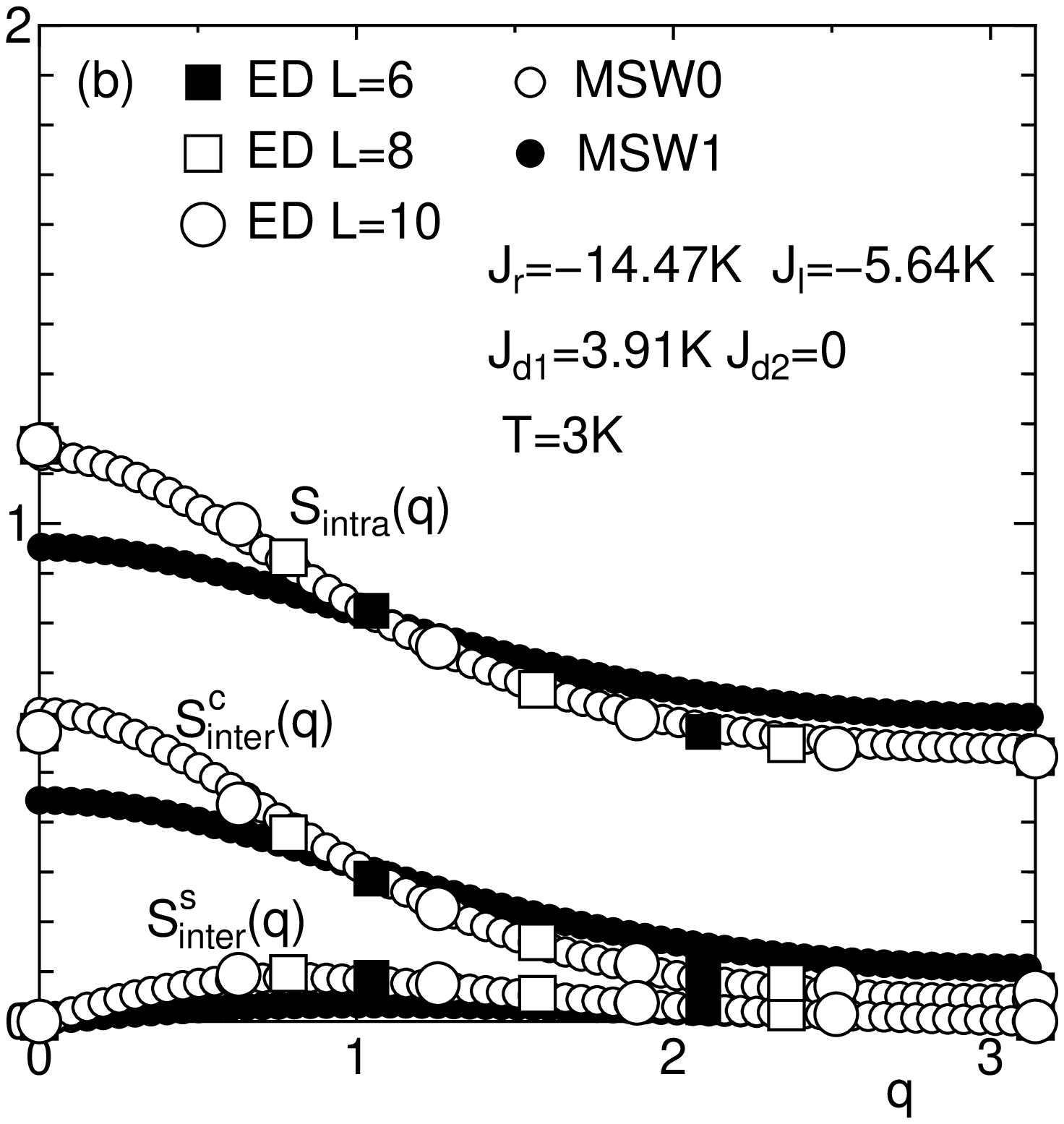}}
\centerline{\includegraphics[width=6cm]{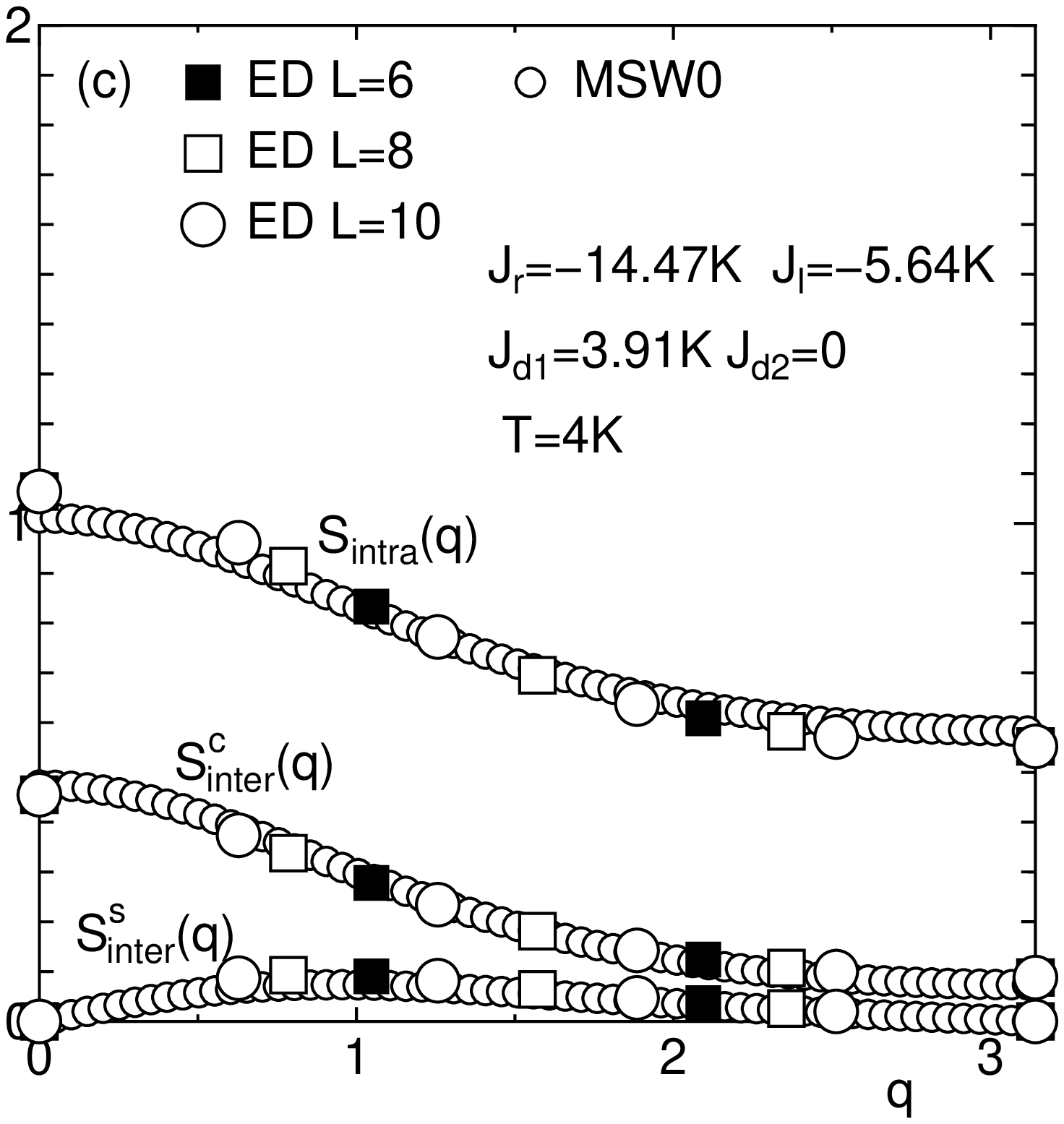}}
\caption{Magnetic structure factors with exchange parameters corresponding to {Cu$^{\rm II}$Cl($O$-$mi$)}$_2$($\mu$-Cl)$_2$ calculated by MSW0 (small open circles) and MSW1 (small filled circles) approximations. The ED data for $L=6$ (big filled squares), 8 (big open squares), and  10 (big open circles)are also shown. Temperatures are (a) 2K, (b) 3K, and (c) 4K. }
\label{sq}
\end{figure}

To confirm that the magnetic short range correlation is appropriately taken into account by the MSW approximations, the magnetic structure factors are shown in Fig. \ref{sq}. With the decrease of temperature, the short range ferromagnetic order develops as expected. The results of ED calculation are also presented in Fig. \ref{sq} for the same choice of exchange constants. Although the MSW1 approximation reproduces the ED results better than MSW0 at $T=2$K, it becomes poor at $T=3$K where the susceptibility also deviates from the ED result. On the other hand, the MSW0 approximation gives reasonable agreement with the ED results even at $T=4$K where the MSW1 equations only have an unphysical solution.

\subsection{Results at the ground state phase transition point}

\begin{figure}
%\centerline{\includegraphics[width=6cm]{frlhik08phase_f_n.eps}\includegraphics[width=6cm]{frlhik05phase_f_n.eps}}
\centerline{\includegraphics[width=6cm]{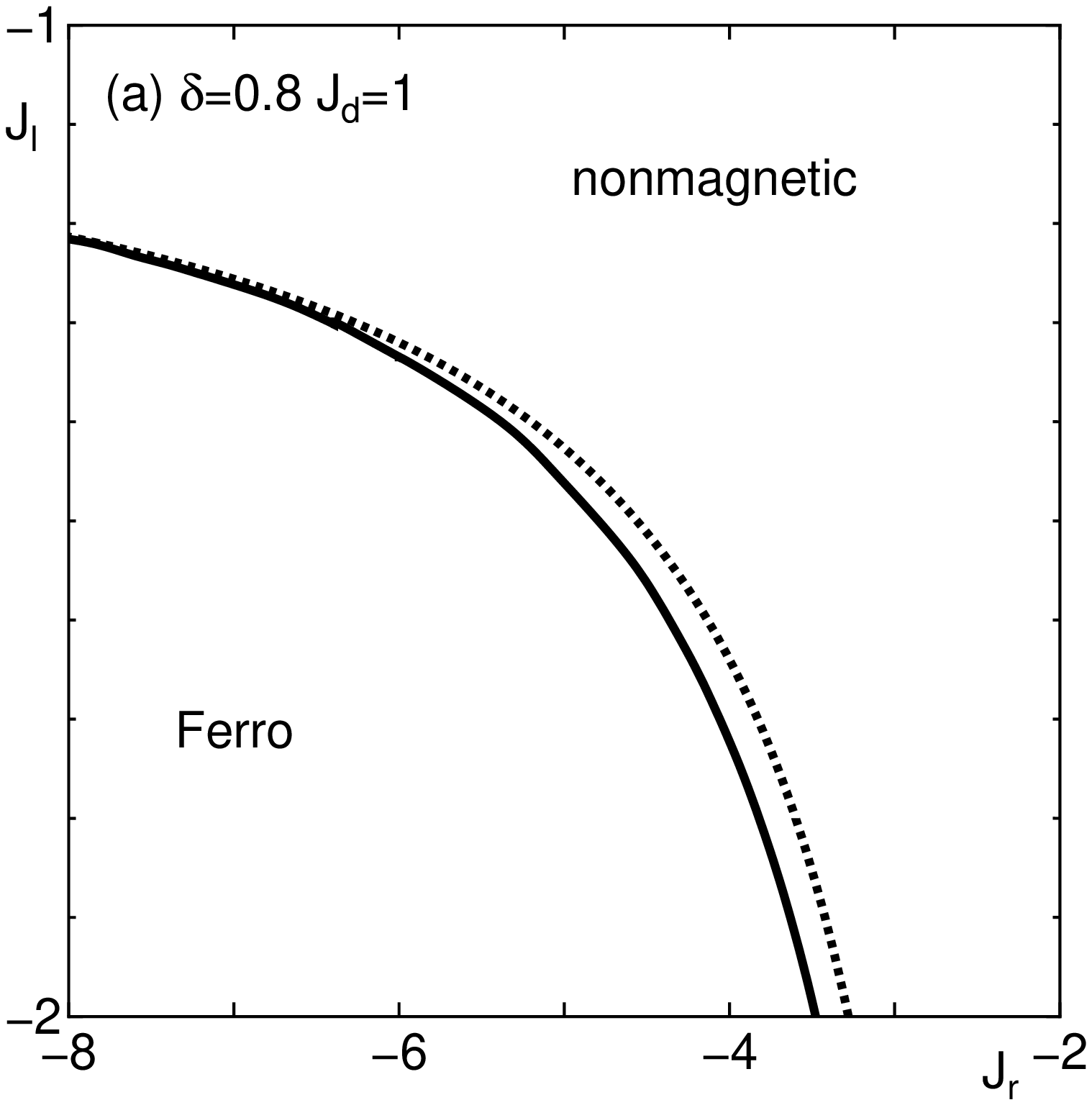}}
\centerline{\includegraphics[width=6cm]{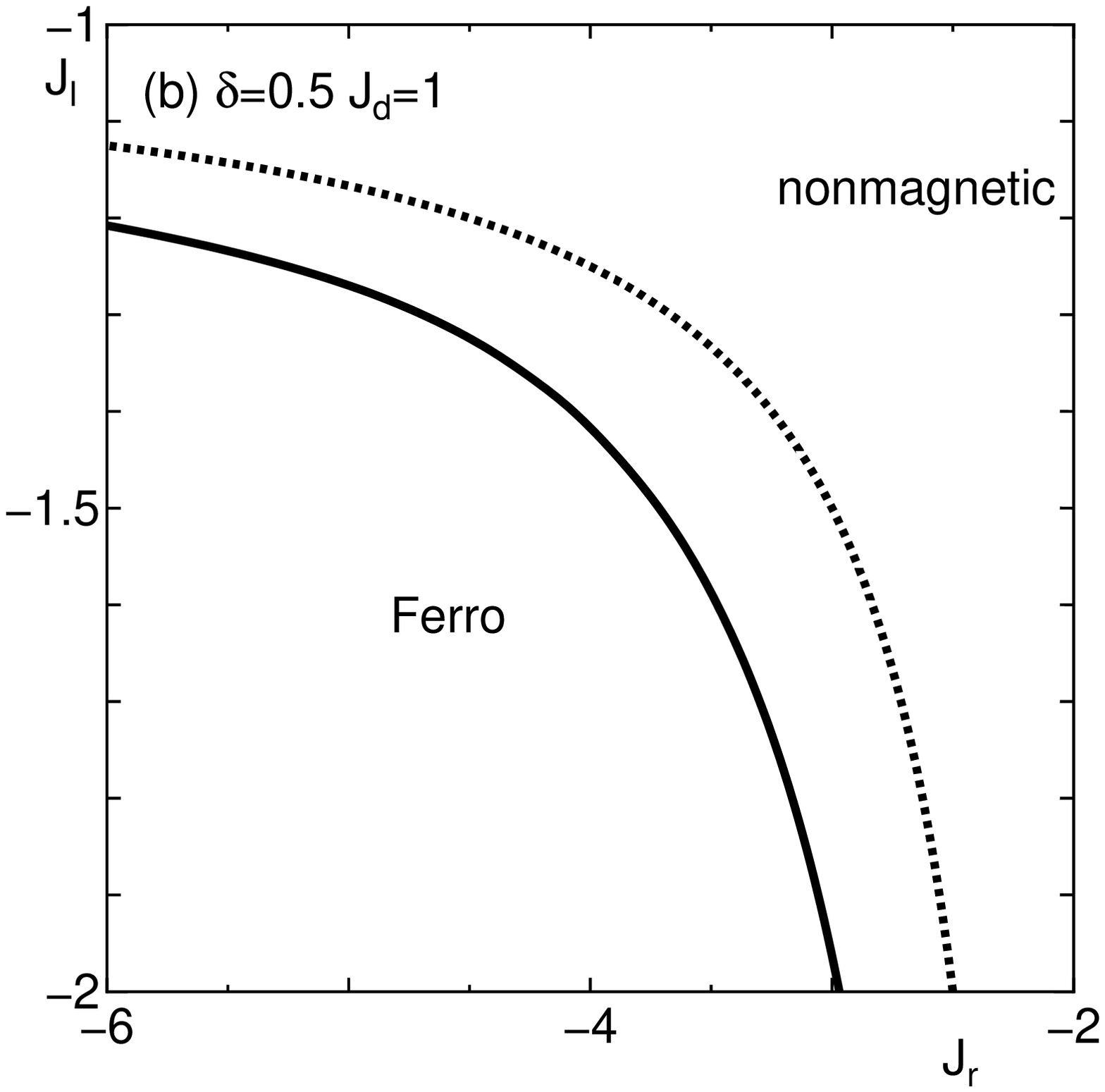}}
\caption{Ground state phase diagram for $\Jd=1$ with (a) $\delta=0.8$ and (b) $\delta=0.5$. The thick line is the ferromagnetic-nonmagnetic transition line and the dotted line is the limit of the stability of the ferromagnetic state given by (\ref{insta}).}
\label{phase}
\end{figure}

\begin{figure}
%\centerline{\includegraphics[width=6cm]{chier80d10dlt08legc_raw_20.eps}}
\centerline{\includegraphics[width=6cm]{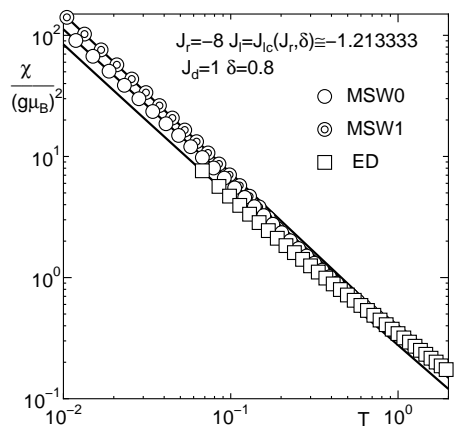}}
\caption{Magnetic susceptibility at the ferromagnetic-nonmagnetic transition point ($\Jr=-8, \delta=0.8, \Jl={\Jlc}(\Jr, \delta) \simeq -1.213333$) which coincides with the limit of the stability of  the ferromagnetic ground state $\Jinst$. The MSW0  and MSW1 results are shown by open circles and double circles, respectively. The extrapolated values from the ED results with $L=6, 8$ and 10 are shown by  open squares. The curves fitted  by $\chi\simeq (A+BT^{1/4})T^{-4/3}$ are also shown. For the fitting of the MSW0 data, $A$ is fixed to  0.1795937 as obtained in Appendix. 
}
\label{chicri}
\end{figure}

With the increase of the frustration, the ferromagnetic ground state becomes unstable against magnon creation at $\Jl=\Jinst$ and the transition to the nonmagnetic ground state takes place. 
However,  the ground state phase transition does not always take place at $\Jl=\Jinst$. The ground state can change from ferromagnetic to  nonmagnetic at $\Jl=\Jlc(\leq\Jinst)$ where the ground state energy of the nonmagnetic state becomes equal to that of the ferromagnetic one. Examples of  ground state phase boundaries and stability limits of the ferromagnetic state are presented  in Fig. \ref{phase}(a) for $\delta=0.8$  and  in Fig. \ref{phase}(b) for $\delta=0.5$.  The ferromagnetic-nonmagnetic phase boundary is determined by the ED method for $L=10$. It is checked that the size dependence is negligible in the scale of this figure.  %At $\delta=1$, the numerical results suggest $\Jlc=\Jinst$. 

First, let us examine the behavior of the magnetic susceptibility at $\Jl=\Jlc$ in the case $\Jlc=\Jinst$. On the line $\Jl=\Jinst$, the excitation energy $\varepsilon_{\alpha}(k)$ is proportional to $k^4$. In this case, the susceptibility calculated by the MSW0 approximation behaves as $T^{-4/3}$ as shown in Appendix.  As an example of the case $\Jlc=\Jinst$, the temperature dependence of the susceptibility  for $\Jr=-8, \Jd=1$ and $\delta=0.8$ and $\Jl=\Jinst\simeq -1.213333$  is shown in Fig. \ref{chicri}. %In this case $\Jinst \simeq \Jlc$ within the numerical accuracy. 
The results calculated by the MSW0 and MSW1 approximations, and those extrapolated from the ED results for $L=6, 8 $ and 10 are shown.  All %Both
 results show the behavior $\chi \sim T^{-4/3}$. However, the ampltude of the susceptibility is overestimated by the MSW approximations. We may understand this discrepancy in the following way: 

Even at $\Jl=\Jinst$, the ferromagnetic state remains one of the ground states. 
In addition, the nonmagnetic state which replace the ferromagnetic states for  $\Jl>\Jinst$ comes into play. However, this state has no magnetic moment and does not contribute to the susceptibility. Also, the low-lying excited states around the nonmagnetic states have small magnetic moments and do not have significant contribution to the susceptibility. Nevertheless, these states have finite statistical weight. Therefore, the contribution from the ferromagnetic state and the excitations around it, which is correctly described by the MSW approximations, can reproduce the leading temperature dependence of magnetic susceptibility, while its actual amplitude is reduced from the results of the MSW approximations. 

\begin{figure}
%\centerline{\includegraphics[width=6cm]{sqr80d10legcra.eps}}
\centerline{\includegraphics[width=6cm]{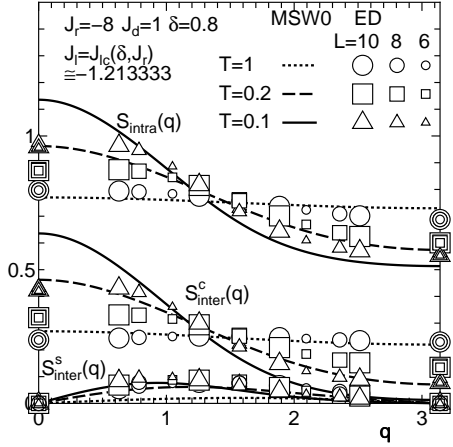}}
\caption{Magnetic structure factors on the ground state phase boundary calculated by MSW0  approximation(lines) and ED(open symbols) methods.  Temperatures are $T=1$ (dotted lines, circles),  0.2 (broken lines, squares) and 0.1 (solid lines, triangles). The big, medium and small symbols represent the system sizes $L=10$, 8 and 6, respectively}
\label{sqcr}
\end{figure}

\begin{figure}
%\centerline{\includegraphics[width=6cm]{sqq0crthrd.eps}}
\centerline{\includegraphics[width=6cm]{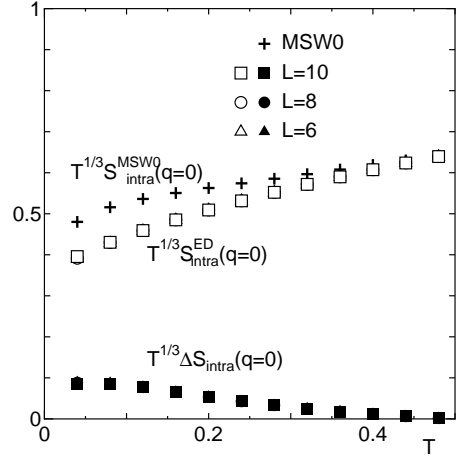}}
\caption{Temperature dependence of $\Sqra(q=0)$  on the ground state phase boundary calculated by the MSW0 approximation (+) and ED methods (open symbols). The difference $\Delta \Sqra (q=0)(=\Sqra^{\rm MSW0}(q=0)-\Sqra^{\rm ED}(q=0))$ between the MSW0 and ED results is also shown (filled symbols). All quantities are multiplied by $T^{1/3}$.}
\label{sqcr0}
\end{figure}

These nonmagnetic states have antiferromagnetic or incommensurate short range correlations induced by  frustration. 
To get more insight into their effects, we calculate the magnetic structure factor for this parameter set as shown in Fig. \ref{sqcr}. It is clearly observed that the $q=0$ component, which is responsible for the susceptibility, is overestimated in the MSW0 approximation. It should be noted that the size dependence of the ED results for $\Sqra(q=0)$ and $\Sqer(q=0)$ is almost negligible. % for the present set of parameters.
 Therefore, this discrepancy is not attributed to the finite size effect. On the other hand, the large $q$ components are underestimated in the MSW0 approximation. Considering the sum rule
\begin{align}
\sum_{q}\Sqra(q)%&=\sum_{q}\frac{1}{L}\sum_{i}\aver{\v{S}_{0,1}\v{S}_{i,1}}\exp(i x_i q)\\
&=\aver{\v{S}_{0,1}\v{S}_{0,1}}=S(S+1),
\end{align}
the enhancement of $q\neq 0$ components suppresses the $q=0$ component for $\Sqra(q)$.
 For $\Sqer(q)$, we have
\begin{align}
\sum_q \Sqer(q)%&=\sum_{q}\frac{1}{L}\sum_{ i}\aver{\v{S}_{0,1}\v{S}_{ i,2}}\exp (i x_i q)\\
&=\aver{\v{S}_{0,1}\v{S}_{ 0,2}}.
\end{align}
In this case, lhs is not a constant. However, this should be close to $1/4$ if $\Jr$ is strongly ferromagnetic as in the present case. Hence, the similar explanation is naturally valid for  $\Sqer(q)$. 
Thus, we may conclude that this discrepancy comes from the frustration induced antiferromagnetic or incommensurate short range order which is not appropriately described by the MSW0 approximation. % Such short range order result from  frustration. 
Fig. \ref{sqcr0} shows the temperature dependence of  $\Sqra(q=0)$ calculated by the MSW0 approximation ($\Sqra^{\rm MSW0}(q=0))$ and that calculated with the ED method ($\Sqra^{\rm ED}(q=0)$). The difference $\Delta \Sqra (q=0)(=\Sqra^{\rm MSW0}(q=0)-\Sqra^{\rm ED}(q=0))$ is also plotted. All quantities are multiplied by $T^{1/3}$. This plot shows that  $T^{1/3}\Delta \Sqra (q=0)$ does not diverge in the low temperature limit. Hence, 
the correction to $\Sqra^{\rm MSW0}(q=0)$ does not diverge with the power stronger than $T^{-1/3}$ %tends to a finite value 
in the low temperature limit. %, if the data for lowest three temperatures with strong size dependence are excluded. 
This means that the power of the leading term of $\chi \sim T^{-4/3}$ is unaffected by this correction.

\begin{figure}
%\centerline{\includegraphics[width=6cm]{chier40d10dlt05leg1st_raw.eps}}
\centerline{\includegraphics[width=6cm]{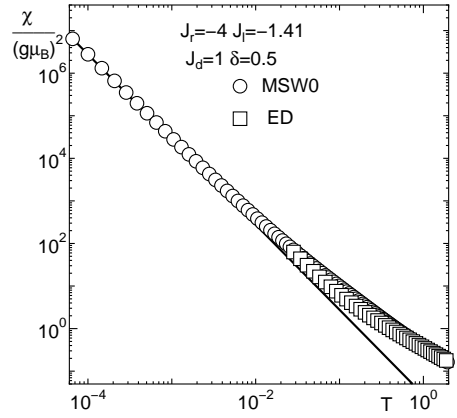}}

\caption{Magnetic susceptibility at the ferromagnetic-nonmagnetic transition point ($\Jr=-4, \Jl=-1.41, \delta=0.5$ ) which is away from the limit of the stability of the ferromagnetic ground state. The MSW0 results are shown by open circles and extrapolated values from the ED results with $L=6, 8$ and 10 are shown by  open squares. The solid line is the fitted line by $\chi\propto{\rm const.}\times T^{-2}$}
\label{chi1st}
\end{figure}

 If the limit of stability of the ferromagnetic phase $\Jinst$ is away from the ferromagnetic-nonmagnetic transition point $\Jlc$,  the susceptibility behaves as $T^{-2}$  as in the case of ferromagnetic ground state. An example is shown in Fig. \ref{chi1st} for $\Jr=-4, \Jd=1$ and $\delta=0.5$ and $\Jl=\Jlc\simeq -1.41$.

\section{Summary and Discussion}

Low temperature properties of the frustrated ferromagnetic ladders with spin-1/2 are investigated using the MSW and ED methods. The magnetic susceptibility and static magnetic structure factors are calculated. %The comparison is made with the ED results. % and experiments with real material. 
It is found that the MSW method gives reasonable agreement with the ED calculation in the intermediate temperature regime below which the ED results show considerable size dependence. On the contrary, the MSW method becomes more reliable in the low temperature regime. Therefore, we conclude that the MSW method is useful  as a complementary method to the ED analysis even in the presence of moderate frustration as far as the ground state remains the ferromagnetic state. 

It is also predicted %shown
 that the MSW %0 
approximation is reliable even at the limit of the stability of  the ferromagnetic ground state  $\Jl=\Jinst$ insofar as the exponent of the leading temperature dependence is concerned.
Although we have explicitly demonstrated $\chi \sim T^{-4/3}$ %exactly
 on the ferromagnetic-nonmagnetic phase boundary, this  behavior should be observed even somewhat away from the transition points at finite temperatures. As the temperature is lowered, the crossover to the true low temperature behavior in ferromagnetic or nonmagnetic phases should take place. Such a crossover behavior can be regarded as a precurser of the ground state phase transition if experimentally observed.

In the present work, we have concentrated on the finite temperature properties in the parameter regime with ferromagnetic ground state. However, our preliminary calculation suggests that the nonmagnetic phase consists of several different exotic phases with and without spontaneous symmetry breakdown. The investigation of these phases will be reported elsewhere.

The authors thank M. Kato and A. Nagasawa for providing their experimental data and for discussion. The numerical diagonalization program is based on the package TITPACK ver.2 coded by H. Nishimori.  The numerical computation in this work has been carried out using the facilities of the Supercomputer Center, Institute for Solid State Physics, University of Tokyo and Supercomputing Division, Information Technology Center, University of Tokyo, and   Yukawa Institute Computer Facility in Kyoto University. This work is supported by a Grant-in-Aid for Scientific Research (C) (21540379) from Japan Society for the Promotion of Science. 

\appendix
\section{}

Let us define  the density of states per site by
\begin{align}
  w(x)&=\frac{1}{2L}\sum _k\left\{\delta (x-\varepsilon _{\alpha }(k))+\delta (x-\varepsilon _{\beta}(k))\right\}.
\end{align}
At low temperatures, only the gapless magnon mode $\alpha$ contributes. We consider the case where the dispersion relation of this gapless mode is given by $\varepsilon_{\alpha}(k)=Ak^n$. The density of states  is then approximated as
\begin{align}
  w(x)&\simeq\frac{1}{2L}\sum _k\delta (x-\varepsilon _{\alpha }(k))\simeq
%\frac{1}{2\pi}\frac{1}{nA(x/A)^{(n-1)/n}}=
\frac{1}{2\pi nA^{\frac{1}{n}}}x^{\frac{1}{n}-1}.
\end{align}
Rewriting \eqref{condition2} and \eqref{chi} using the density of states, we have
\begin{align}
  S=&\int _0^{\infty }\frac{w(x)dx}{\exp (xT^{-1}+v)-1} =\frac{T^{\frac{1}{n}}}{2\pi nA^{\frac{1}{n}}}F\left(\frac{1}{n} ,v\right)\Gamma\left(\frac{1}{n}\right)\label{S}, \\
  \chi =&\frac{(g\mu _{\text{B}})^2}{3T}\int _0^{\infty }\frac{\exp (xT^{-1}+v)w(x)dx}{(\exp (xT^{-1}+v)-1)^2} \notag\\
&=\frac{(g\mu _{\text{B}})^2}{3}\frac{T^{\frac{1}{n}-1}}{2\pi nA^{\frac{1}{n}}}\Gamma\left(\frac{1}{n} \right)F\left(\frac{1}{n}-1,v\right),
\label{kai}
\end{align}
where we define
\begin{align}
  v\equiv -\mu /T
\end{align}
and $F(\alpha,v)$ is the Bose-Einstein integral function defined by
\begin{align}
 F(\alpha ,v)=\frac{1}{\Gamma (\alpha )}\int _0^{\infty }\frac{u^{\alpha -1}du}{e^{u+v}-1}.
\end{align}
The behavior of this function for small $v$ is known \cite{RobinsonPhysRev.83.678}. For the present purpose, we only need the formula for noninteger $\alpha $.
\begin{align}
  F(\alpha ,v)=\Gamma (1-\alpha )v^{\alpha -1}+\sum _{l=0}^{\infty }(l!)^{-1}(-v)^l\zeta (\alpha -l).
\end{align}
For the calculation of the susceptibility, the partially integrated form
\begin{align}
  \int _0^{\infty }\frac{u^{\alpha -1}e^{u+v}}{(e^{u+v}-1)^{2}}du=\Gamma (\alpha )F(\alpha -1,v)
\end{align}
is useful. 
The equations (\ref{S}) and (\ref{kai}) can be expressed as
\begin{align}
  S\simeq&\frac{T^{\frac{1}{n}}}{2\pi nA^{\frac{1}{n}}}\Gamma\left(\frac{1}{n}\right)\left\{\Gamma \left(1-\frac{1}{n}\right)v^{\frac{1}{n} -1}+\zeta\left(\frac{1}{n}\right)+O(v^1)\right\},\label{app:s}
\\
  \chi \simeq&\frac{(g\mu _{\text{B}})^2}{3}\frac{T^{\frac{1}{n}-1}}{2\pi nA^{\frac{1}{n}}}\Gamma\left(\frac{1}{n}\right)\notag\\
&\times\left\{\Gamma \left(2-\frac{1}{n}\right)v^{\frac{1}{n}-2}+\zeta \left(\frac{1}{n} -1\right)+O(v^1)\right\}.\label{app:chi}
\end{align}
Solving (\ref{app:s}) with respect to  $v$ and substituting into (\ref{app:chi}), we find
\begin{align}
 \chi  \simeq&\frac{(g\mu _{\text{B}})^2S}{3}\frac{n-1}{n}
\left({2 nSA^{\frac{1}{n}}\sin\frac{\pi}{n}}\right)^{\frac{n}{n-1}}T^{-\frac{n}{(n-1)}}\nonumber\\
&\times\left\{1+\frac{1-2n}{n-1}\zeta\left(\frac{1}{n}\right)\Gamma\left(\frac{1}{n}\right)\frac{T^{\frac{1}{n}}}{2\pi nSA^{\frac{1}{n}}}\right.\notag\\
&\left.-\frac{n(1-2n)}{2(n-1)^2}\zeta^2\left(\frac{1}{n}\right)\Gamma^2\left(\frac{1}{n}\right)\left(\frac{1}{2\pi nSA^{\frac{1}{n}}}\right)^2T^{\frac{2}{n}}\right\}.
\end{align}
Within the ferromagnetic phase,  $n=2$. Setting $A=\mathscr{J}S$, we have
\begin{align}
 \chi \simeq &\frac{8(g\mu _{\text{B}})^2S^4}{3T^2}\mathscr{J}\left\{1-\frac{3T^{\frac{1}{2}}\zeta\left(\frac{1}{2}\right)}{4S\sqrt{\pi \mathscr{J}S}}+\frac{3\zeta^2 \left(\frac{1}{2}\right)}{16\pi S^2\mathscr{J}S}T\right\}.
\end{align}
At the stability limit of the ferromagnetic phase,  $n=4$. Hence, we have
\begin{align}
 \chi   \simeq&\frac{{(g\mu_{\text{B}})^2}S^{\frac{7}{3}}A^{\frac{1}{3}}2^{\frac{4}{3}}}{T^{\frac{4}{3}}}
\left\{1-\frac{7}{3}\frac{\zeta\left(\frac{1}{4}\right)\Gamma \left(\frac{1}{4}\right)T^{\frac{1}{4}}}{8\pi SA^{\frac{1}{4}}}\right.\notag\\
&\left.+\frac{7}{9}\left(\frac{\zeta \left(\frac{1}{4}\right)\Gamma \left(\frac{1}{4}\right)}{8\pi SA^{\frac{1}{4}}}\right)^2T^{\frac{1}{2}}\right\}.
\end{align}
In the low temperature limit, we have
\begin{align}
\lim_{T\rightarrow 0}\frac{\chi T^{4/3}}{(g\mu_{\text{B}})^2}
&=\frac{A^{\frac{1}{3}}}{2}
\end{align}
for $S=1/2$. Numerically $A \simeq 0.04634074$ for $\Jr=-8, \Jd=1, \delta=0.8, \Jl=\Jlc$, $\dfrac{\chi T^{4/3}}{(g\mu_{\text{B}})^2}\simeq 0.1795937$. This value is used in the fitting of the MSW0 data in %plotted by a filled circle in the inset of 
Fig. \ref{chicri}.
%\begin{thebibliography}{50}

%\bibliography{frulad_msw.1.bib}

\begin{thebibliography}{10}

\bibitem{kato.ejic10}
M.~Kato, K.~Hida, T.~Fujihara, and A.~Nagasawa: Eur. J. Inorg. Chem. {\bfseries
  2011} (2011) 495.

\bibitem{Barnes.PRB93}
T.~Barnes, E.~Dagotto, J.~Riera, and E.~S. Swanson: Phys. Rev. B {\bfseries 47}
  (1993) 3196.

\bibitem{Gopalan.PRB94}
S.~Gopalan, T.~M. Rice, and M.~Sigrist: Phys. Rev. B {\bfseries 49} (1994)
  8901.

\bibitem{hik.prb10}
T.~Hikihara and O.~A. Starykh: Phys. Rev. B {\bfseries 81} (2010) 064432.

\bibitem{HakobyanPRB01}
T.~Hakobyan, J.~H. Hetherington, and M.~Roger: Phys. Rev. B {\bfseries 63}
  (2001) 144433.

\bibitem{taka.ptp86}
M.~Takahashi: Prog. Theor. Phys. Suppl. {\bfseries 87} (1986) 233.

\bibitem{Takahashi.PRL87}
M.~Takahashi: Phys. Rev. Lett. {\bfseries 58} (1987) 168.

\bibitem{Takahashi.PRB90}
M.~Takahashi: Phys. Rev. B {\bfseries 42} (1990) 766.

\bibitem{Takahashi.PRB89}
M.~Takahashi: Phys. Rev. B {\bfseries 40} (1989) 2494.

\bibitem{Ohara-YosidaJPSJ89}
K.~Ohara and K.~Yosida: J. Phys. Soc. Jpn. {\bfseries 58} (1989) 2521.

\bibitem{Ohara-YosidaJPSJ90}
K.~Ohara and K.~Yosida: J. Phys. Soc. Jpn. {\bfseries 59} (1990) 3340.

\bibitem{Mermin.PRL66}
N.~D. Mermin and H.~Wagner: Phys. Rev. Lett. {\bfseries 17} (1966) 1133.

\bibitem{HerzogPRB11}
A.~Herzog, P.~Horsch, A.~M. Ole\ifmmode~\acute{s}\else \'{s}\fi{}, and
  J.~Sirker: Phys. Rev. B {\bfseries 84} (2011) 134428.

\bibitem{Shanks}
D.~Shanks: J. Math. Phys. {\bfseries 34} (1955) 1.

\bibitem{RobinsonPhysRev.83.678}
J.~E. Robinson: Phys. Rev. {\bfseries 83} (1951) 678.

\end{thebibliography}

\end{document}